# Direct ultrashort pulse generation by intracavity nonlinear compression

**L. M. Zhao, D. Y. Tang, T. H. Cheng**

School of Electrical and Electronic Engineering, Nanyang Technological University, Singapore

**C. Lu**

Department of Electronic and Information Engineering, Hong Kong Polytechnic University

Direct generation of ultrashort, transform-limited pulses in a laser resonator is observed theoretically and experimentally. This constitutes a new type of ultrashort pulse generation in mode-locked lasers: in contrast to the well-known solitons (hyperbolic secant like), dispersion-managed solitons (Gaussian-like), and parabolic pulses plus external compression, ultrashort pulse solutions to the nonlinear wave equations that describe pulse evolution in the laser cavity are observed. Stable ultrashort, transform-limited pulses exist with optical spectrum broader than the gain bandwidth of the amplifier, and this has practical application for other lasers.

Ultrashort pulse generation has been extensively studied since the ultrashort pulses are very useful for the generation of many widely differing nonlinear physical phenomena, including stimulated Raman effect, Brillouin-scattering process, parametric four-wave mixing, and self-phase modulation [1]. In the field of optics, there are two main setups to achieve ultrashort pulse generation, solid-state lasers and fiber lasers. Although generally fiber lasers cannot achieve performance better than solid-state lasers, they still attract many scientists since fiber lasers are more compact in size, cost effective and easier to operate. Recently, fiber lasers that can offer performance comparable with that of solid state lasers with nearly all the advantages of stability and compactness of all-fiber lasers are reported [2]. Ultrashort pulses generated in fiber lasers sometimes are recognized as solitons [3]. Solitons in fiber lasers are generated due to the balanced interaction of dispersive and nonlinear effects after mode locking [1]. However, the peak power of solitons generated in fiber lasers is limited by soliton instability. Increasing laser gain can only increase the number of solitons in the laser rather than their peak power. In order to achieve ultrashort pulse with higher peak power, so-called stretched-pulse mode-locking technique was introduced [4,5]. Instead of constructing the laser cavity by using entirely the fibers of anomalous group velocity dispersion (GVD), fiber of normal GVD is also incorporated. Such lasers were also known as dispersion-managed lasers [6]. As the optical pulse circulating in such a laser cavity is temporally stretched and compressed in one round trip, the average peak power of the mode-locked pulse becomes lower than that of a transform-limited pulse of the same spectral bandwidth. Superfluous nonlinearity is avoided and pulse with larger energy can be obtained. The basic idea of dispersion-managed fiber lasers is to possibly reduce the nonlinearity of the formed pulse. When the net cavity dispersion is anomalous and big, hyperbolic secant like pulses can still be obtained; when the net cavity dispersion is close to zero, dispersion-managed solitons (DMSs) [7] with Gaussian-like pulse profile are generated, where the output pulses have picosecond duration and are linearly chirped. External pulse compression system can dechirp the pulses and obtain sub-100fs pulses. It has been demonstrated that in a stretched pulse fiber laser pulses of shortest pulsewidth and highest energy are obtained in the net positive dispersion condition with externally optimal compression [8].

Parabolic pulse generation combined with external pulse compression system [9,10] is another fundamental method to achieve high power ultrashort pulses, where in high gain optical fiber amplifiers with normal dispersion a linearly chirped parabolic pulse with wave-breaking free evolution is generated. Such self-similar pulses can bear strong nonlinearity without wave breaking comparing with solitons. However, self-similar propagation of intense pulses would be disrupted if the pulse encounters any limitation to its spectral bandwidth [11]. Gain-guided solitons could be generated in fiber lasers of normal dispersion [12,13]. In contrast to the parabolic pulses, gain-guided solitons also have large pulse energy but nonlinear chirp. Therefore, it is possible to obtain ultrashort pulses by using the nonlinear pulse compression to dechirp the nonlinear pulse chirp accumulated during the nonlinear pulse amplification process.

Solitons and DMSs are both eigen solutions of nonlinear wave equations while parabolic pulses are asymptotic solutions to the governing equations. They need a mechanism to compensate any changes and hence repeat themselves after one round trip in a cavity. In a laser resonator [10] made of major fiber segments with normal GVD, the self-similar parabolic pulses are generated by special design where the pulse can be amplified with negligible GVD and nonlinearity, and the linear chirp accumulated in the desired self-similar propagation is compensated by a dispersive delay line that provides anomalous GVD with negligible nonlinearity. The mechanisms of the generation of DMSs and parabolic pulse in a sense are both try to eliminate nonlinear phase shifts while accommodating more pulse energy in contrast to solitons. In another word, they are "linear" cases for pulse generation. Actually, due to the intrinsic nonlinear properties of fibers there exists a general "nonlinear" case that is based on the nonlinear pulse amplification and compression technique to achieve high peak power ultrashort pulses directly from a fiber laser. Soliton is a representative nonlinear case without compression in the negative dispersion regime. In the dispersion-managed laser resonator comprising of normal dispersive gain media, there exists another nonlinear case that can result in direct generation of ultrashort, transform-limited pulses.

Here we report numerical predictions of the existence of stable pulses that is transform-limited with broader bandwidth comparing with the gain bandwidth of the amplifier in a laser cavity, along with clear experimental evidence of direct generation of ultrashort, transform-limited pulses in a fiber lasers. This method constitutes a new type of ultrashort pulse generation in a mode-locked laser: qualitatively distinct from the generation of well-known solitons (hyperbolic secant like) [3], DMSs (Gaussian-like) [7], and parabolic pulses plus external compression [9,10]. The mechanism of this new method and the features of the generated pulses are presented.

Pulse formation in a femtosecond laser is typically dominated by the interplay between dispersion and nonlinearity [14]. The basic features of the laser can be described by the extended Ginzburg-Landau equation, which takes into account the interplay among the cavity dispersion, nonlinear self-phase modulation and properties of the laser gain medium [15].

Soliton fiber lasers are limited to low pulse energy and small peak power. In a soliton fiber laser, the nonlinear phase shift of a mode-locked pulse accumulated within one cavity round-trip is easy to exceed a certain value with higher pump power, which would generate multiple soliton operation. DMSs can tolerate higher nonlinear phase shifts than conventional solitons, consequently they have much larger pulse energy. However, the peak clamping and multiple pulse generation, noise-like pulse formation [16] still exist when higher pump power is available in the laser. Self-similar pulses can tolerate much higher pulse energy comparing with DMSs since it can remain parabolic but expanding pulse width after propagation through fibers with normal dispersion as long as there is no bandwidth restriction.

Same as conventional solitons, DMSs, and parabolic pulses, direct transform-limited pulses are another solutions to the governing nonlinear wave equations. Different from soliton generation, the pulses are amplified in an amplifier of normal dispersion. Furthermore, it is gain-guided after its propagation along the gain media with normal dispersion, which is different from parabolic pulses. It is natural to use the intrinsic nonlinear pulse compression intracavity since the pulse has nonlinear chirp when it is gain-guided amplified in the amplifier. Therefore, it is possible to achieve direct transform-limited pulse output.

The fiber laser is illustrated in Fig. 1. The gain-guided amplification is provided by a segment of EDF with normal GVD. After amplification, the pulse is nonlinearly dechirped by an optimized segment of single-mode fiber (SMF) with negative GVD. The transform-limited pulses are output by a beam splitter, which also serves as the polarizer, accompanying with other polarization controllers, to mode lock the pulses. Another segment of SMF is incorporated to make the net cavity GVD is positive and close to zero, where the narrowest pulse with large pulse energy can be generated. Since the pulses are nonlinear amplified and nonlinear compressed, no special attention is needed except that the gain medium has normal dispersion to provide gain-guided amplification and the length of the SMF following the EDF is optimized for the direct output of transform-limited ultrashort pulses.

Numerical simulations basing on the laser were carried out to verify the ultrashort pulse solutions. The parameters of the numerical model correspond to the experimental setup. A segment of EDF 210 cm in length is used for the gain-guided amplification, followed with a segment of SMF 62cm in length to nonlinear compress the pulses. The length of the fiber is carefully selected so that the output pulse by the beam splitter is near transform-limited. Pulse propagation in each segment is modeled by

$$\begin{cases} \frac{\partial u}{\partial z} = i\beta u - \delta\frac{\partial u}{\partial t} - \frac{ik''}{2}\frac{\partial^2 u}{\partial t^2} + \frac{ik'''}{6}\frac{\partial^3 u}{\partial t^3} + i\gamma(|u|^2 + \frac{2}{3}|v|^2)u + \frac{i\gamma}{3}v^2u^* + \frac{g}{2}u + \frac{g}{2\Omega_g}\frac{\partial^2 u}{\partial t^2} \\ \frac{\partial v}{\partial z} = -i\beta v + \delta\frac{\partial v}{\partial t} - \frac{ik''}{2}\frac{\partial^2 v}{\partial t^2} + \frac{ik'''}{6}\frac{\partial^3 v}{\partial t^3} + i\gamma(|v|^2 + \frac{2}{3}|u|^2)v + \frac{i\gamma}{3}u^2v^* + \frac{g}{2}v + \frac{g}{2\Omega_g}\frac{\partial^2 v}{\partial t^2} \end{cases} \quad (1)$$

where u and v are the normalized envelopes of the optical pulses along the two orthogonal polarized modes of the optical fiber. $2\beta = 2\pi\Delta n/\lambda$ is the wave-number difference between the two modes. $2\delta = 2\beta\lambda/2\pi c$ is the inverse group velocity difference. $k''$ is the second order dispersion coefficient, $k'''$ is the third order dispersion coefficient and $\gamma$ represents the nonlinearity of the fiber. g is the saturable gain coefficient of the fiber and $\Omega_g$ is the bandwidth of the laser gain. For undoped fiber $g = 0$; For erbium doped fiber, we have considered its gain saturation as

$$g = G\exp[-\frac{\int(|u|^2 + |v|^2)dt}{E_{sat}}] \quad (2)$$

where G is the small signal gain coefficient and $E_{sat}$ is the normalized saturation energy.

In accordance with our experimental condition we have used the following parameters for the simulations: $\gamma$=3 $W^{-1}km^{-1}$, $k''_{EDF}$=32 (ps/nm)/km, $k''_{SMF}$= -20 (ps/nm)/km, $k'''$=0.1 ($ps^2$/nm)/km, $\Omega_g$ = 24 nm, gain saturation energy Esat=300 pJ, cavity length L=$2.4_{SMF}$+$2.1_{EDF}$+$0.6_{SMF}$=5.1 m, beat length $L_b$=L/2, and the orientation of the intracavity polarizer to the fiber fast birefringent axis $\Psi$= 0.152π.

We used the standard symmetric split-step method to solve the numerical model. Pulse tracing technique is use to track the pulse evolution in the laser cavity. The simulations run until the total output energy tends to a constant value independent of the round-trip numbers. The numerical accuracy is verified by checking that the results are fixed after doubling the sampling resolution [10]. Fig. 2 shows a typical solution obtained with the cavity linear phase delay bias setting [16] being 1.85π and G=2500 $km^{-1}$. Ultrashort pulse with nearly 5 kW peak power is obtained. The pulsewidth of the generated pulse is about 53.4 fs and the 3-dB bandwidth is about 75.6 nm. Stable output pulse train and intracavity pulse evolution are shown in Fig. 3. The small side pulses that accompany the main ultrashort peak are caused by the imperfectly dechirp process, which also results in the slightly bigger value of time-bandwidth product of the generated ultrashort pulse. The pulse is fast nonlinear compressed after it is gain-guided amplified as shown in Fig. 3b, the rejected polarization component of the optimized ultrashort pulse with high peak power is then output. Remained pulse energy continues to propagate in the cavity.

The parameters of the laser and the operation conditions are specially chosen. First, the pulse needs to be operated with the gain-guided amplification so that the pulse could be nonlinearly amplified. Secondly, the length of the SMF following the EDF should be optimized to nonlinear compress the pulse and generate the transform-limited pulse. Thirdly, the length of the SMF segment after the output beam splitter needs to be appropriately selected so that the overall cavity dispersion is suitable: too short SMF would make the total dispersion so large that only gain-guided solitons could be generated [13]; too long SMF would make the laser being operated in the soliton regime. Numerically we found that the cavity-induced peak power clamping effect [16] still exists but only for large linear cavity phase delay bias setting, therefore the peak power of the generated pulse is operation dependent. Further increasing the cavity linear phase delay bias setting, higher peak power could be achieved.

Encouraged by the simulation results, we set up an EDF laser with the parameters listed above. Self-started mode locked operation is achieved by the nonlinear polarization rotation (NPR) technique. A polarization-dependent beam splitter is used as the output coupler, which outputs only the rejected polarization of the NPR mode locking. Therefore, the output coupling strength depends on the orientation of the waveplates, and even with large output coupling self-started mode-locking can still be achieved. The laser mode locks easily provided the wave plates are appropriately set and pump power is high enough. However, the mode-locked pulse is generally gain-guided soliton with steep spectral edges and linear chirp, which could be dechirped by an external segment of SMF. Carefully tuning the wave plates into the large cavity linear phase delay bias setting could generate directly output of ultrashort pulses. Results obtained with cavity dispersion of 0.009 $ps^2$ are presented as shown in Fig. 4.

Direct output of pulse, with 46.2 fs pulse width (hyperbolic secant profile assumed), 63.2 nm 3-dB bandwidth is achieved with a pump power of about 700 mW. The time-bandwidth-product is 0.365, slightly larger than that of the transform-limited pulse. The peak power is about 52.1 kW if we omit the small side pulses, at an average power of 92.5 mW. However, the side pulse has long tail, which could share up to about 50% of the whole pulse energy. Therefore the transient peak power of the central pulse may be reduced to about 25 kW. The repetition rate is 38.5 MHz. The pulse profile and its optical spectrum distinguish from conventional solitons, DMSs, parabolic pulses, and gain-guided solitons. It is note that although the laser cavity has a similar configuration as that of the stretched pulse lasers, the mechanism of the ultrashort pulse generation is completely different. In contrast to the stretched pulse lasers where dispersion management is use to reduce the effective cavity nonlinearity, in the current laser resonator the normal dispersive gain media is used to provide gain-guided amplification that could limit multiple pulse operation, and the negative dispersive fiber segments (SMF) are chosen to optimally nonlinear compress the amplified pulse and modify the total cavity dispersion to bypass the soliton and gain-guided soliton operation regime.

In this Letter we have focused on fiber lasers, the critical requirements for the direct generation of transform-limited pulse with high peak power are: 1) nonlinear amplification (gain-guided amplification) of propagating pulses and 2) SMFs with appropriate length both satisfy the direct output of transform-limited pulse and exclusive operation regime. Therefore, there are no limitations on the operation laser type. Provided that the laser could fulfill above requirement, direct output of transform-limited pulses with high peak power in solid-state lasers are possible [17].

In conclusion, we have observed direct generation of ultrashort transform-limited pulse with high peak power in a laser resonator. This new type of ultrashort pulse generation is qualitatively distinct from well-known conventional solitons, DMSs, and parabolic pulse plus external compression system. Transform-limited pulses with broader bandwidth compared with gain media in the laser cavity exist under the optimized compression of the gain-guided amplified pulses. The pulse operation regime is independent of other types of ultrashort pulses formation. It may overcome most problems associated with the conventional soliton fiber lasers: low peak power for solitons, extra pulse compression system for DMSs and parabolic pulses, and broad pulse width for gain-guided solitons.

## Figure Captions:

Fig. 1 Experimental setup used for direct generation of transform-limited pulses.

Fig. 2 Numerical simulation results. (a) Temporal intensity profile; (b) Optical spectrum.

Fig. 3 Numerical simulation results. (a) Stable ultrashort pulse train; (b) Pulse evolution along the cavity.

Fig. 4 Experimental results. (a) Autocorrelation trace; (b) Optical spectrum.

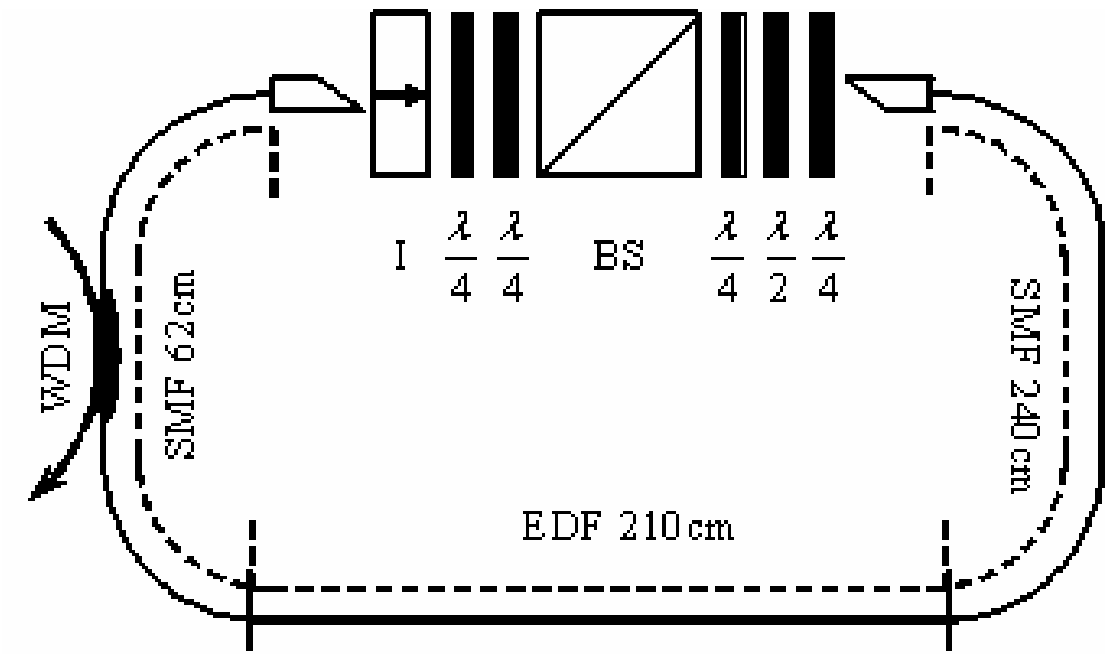

Fig. 1 D. Y. Tang et al.
"Direct ultrashort pulse generation by intracavity nonlinear compression"

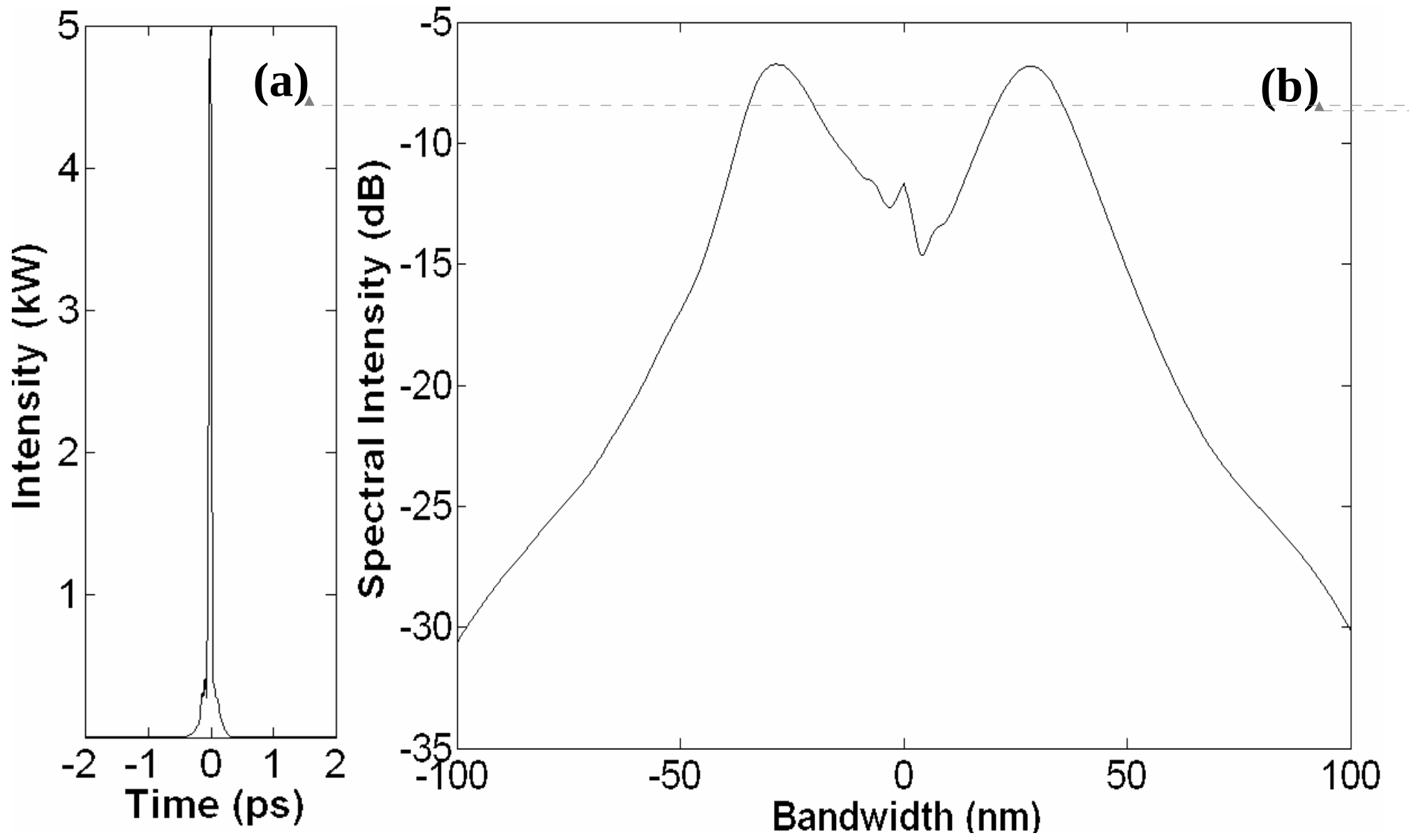

Formatted: Font: 16 pt

Formatted: Font: 16 pt

Fig. 2 D. Y. Tang et al.
"Direct ultrashort pulse generation by intracavity nonlinear compression"

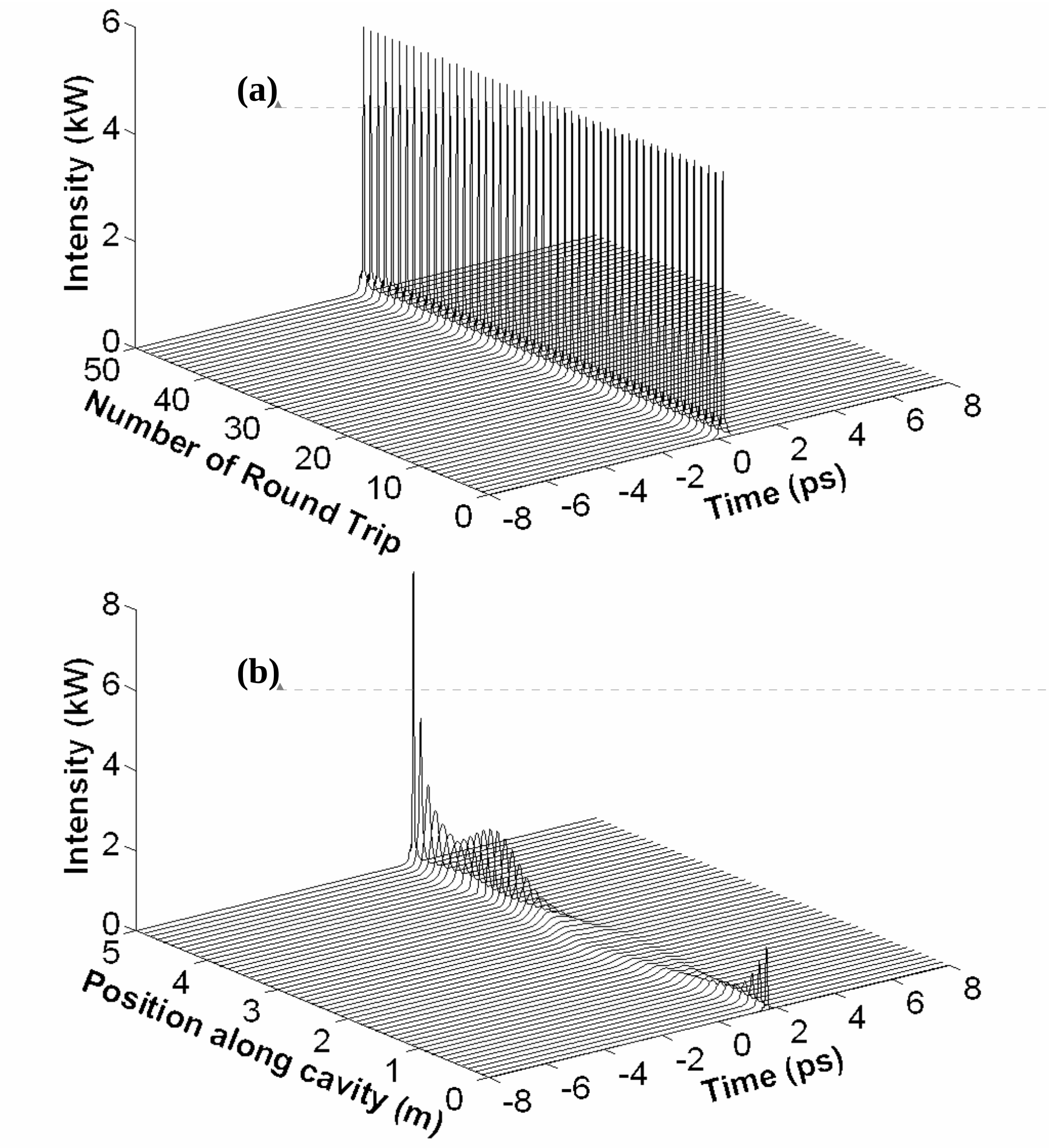

**Formatted:** Font: 16 pt

**Formatted:** Font: 16 pt

Fig. 3 D. Y. Tang et al.
"Direct ultrashort pulse generation by intracavity nonlinear compression"

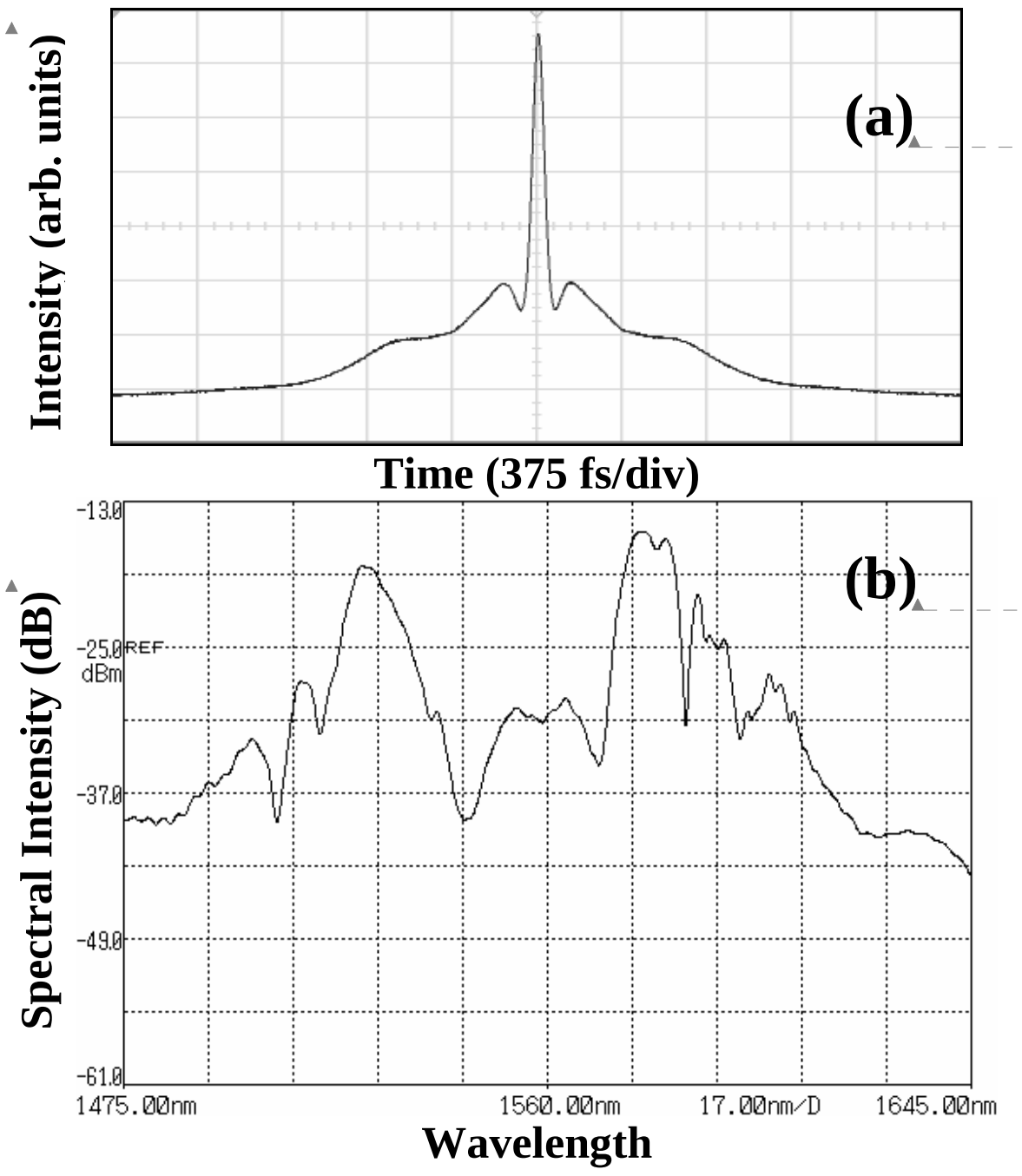

Formatted: Font: 10 pt
Formatted: Font: 16 pt
Formatted: Font: 16 pt
Formatted: Font: 10 pt

Fig. 4 D. Y. Tang et al.
“Direct ultrashort pulse generation by intracavity nonlinear compression”